\documentclass[seceq]{ptptex}
\newcommand{\vect}[1]{\mbox{\boldmath$#1$}}

\renewcommand*{\theequation}{\thesection.\arabic{equation}}
\usepackage{graphicx}
\title{Origin of the Strong Toroidal Magnetic Field in Magnetars}

\author{Naoki \textsc{Onishi}$^{1}$, Tomoyuki \textsc{Maruyama}$^{2}$,
}
\inst{
$^{1}$Department of Physics, Tokyo Metropolitan University,\\ Minamiohsawa, Hachioji
192-0397, Japan 

$^{2}$College of Bioresource Sciences,
Nihon University, \\ Kameino 1866.
Fujisawa 252-8510, Japan
}

\abst{
We propose a simple model of chiral asymmetry to interpret the possible origin of the strong toroidal magnetic field
$\approx 1 \times 10^{12} $ T suggested by analysis of the hard X-ray detector data. 
In a typical size magnetar, the electron is a quantum degenerate system exhibiting
non-local feature due to total anti-symmetrization in the entire system.
Since Fermi energy is of the order of 100 Mev, the electrons relevant to dynamics forming
the strong toroidal magnetic field are ultra-relativistic.
In rotating neutron star, the electron system shares the angular momentum with the hadron
and forms a circular flux of an orbiting current. 
During gravity collapse, high energy electrons are captured by protons converted to neutrons
through parity violating weak interaction, while positive chirality electrons become remnant
resulting in chiral asymmetry.
It is demonstrated that the relevant orbiting electrons having positive chirality are
predominantly in positive helicity states resulting in spin like current of the Gordon decomposition of
Dirac current described by Clifford number and that the current gives rise to strong toroidal magnetic field.
The interaction between the spin like current and the vector potential of the toroidal magnetic field
sustains the helicity and the chirality.
}

\begin{document}
\maketitle
 
\section{Introduction}\label{intro}
Neutron stars (NS) observed as a pulsar are rotating with extremely regular spin period $P$
increasing also very regularly.
The spinning down phenomenon is attributed to magnetic dipole radiation.
The rotating magnetic dipole model estimates surface magnetic field (MF) by
$B_{\rm s}=3.2 \times 10^{3} \sqrt{P\dot{P}}$ TT (1 Tera Tesla $ = 10^{16} $ Gauss) with time derivative of the period $\dot{P}$\cite{HL06}.
In a more direct manner, the MF is estimated from the energy spectrum of
X-ray, according to the electron cyclotron resonance scattering feature.
Neutron stars having strong magnetic field, e.g. higher than the quantum critical field $B_{\rm QED}=4.414 \times 10^{-3}$TT is referred as magnetar \cite{DT92}.
 
The periodgram analysis of the hard x-ray detector data suggests distributions of MF in a magnetar having an inner structure of a toroidal and the strength of $B_{\rm t} (\approx 1{\rm TT})$,  
much stronger than a poloidal (or dipole) field $B_{\rm d} (0.013 {\rm TT})$\cite{MEH14}.
This suggestion is made in terms of deformation caused by the magnetic pressure, and therefore the strong toroidal MF is difficult to observe directly.
It is worthwhile to study theoretically the possible origin of such MF. 

The toroidal MF was theoretically discussed under the consideration of the chiral asymmetry of electrons produced through the parity-violating weak interaction process during core collapse of supernovae by A. Ohnishi and N. Yamamoto\cite{OY14}. They relied on somewhat mathematical concept of magnetic helicity being proportional
to the Gauss linking number and approximate conserved number to explain the long sustainability of such MF.
However, they considered MF formed in the plasma or magnetohydrodynamic matter
and did not show its structure or mechanism of formation.
It was criticized in terms of the quick decay due to chiral mixing through mass term\cite{GKR14}.
In the present study following the idea of chiral asymmetry brought about in the process of gravity collapse,
we provide a clear explanation by demonstrating a more practical structure and dynamics in terms of the spin current in Gordon decomposition\cite{Go28d} of Dirac current described by Clifford number \cite{Di28}.
Our major issues is that the quantum degeneracy of electron system exhibits non-local feature due to total anti-symmetrization in the entire NS system.
In other words the long range kinematical correlation defines the global structure of its current and MF.
 
In order to figure out a general feature of the system, let us take up an example of a NS whose mass and radius are $1.5M_{\odot} $and 12 km, respectively.
The total number of baryon is estimated to be $1800 N_{\star}$, where the big number $N_{\star}$ stands for $10^{54}=(10^{18})^{3}$ based on scaling factor $1 {\rm km}=10^{18} {\rm fm}$.
The average baryon density turns out to be $0.25 N_{\star} {\rm km}^{-3} \!\! =0.25 {\rm fm}^{-3}$, which is 1.4 times higher than the one of an atomic nucleus.
In this density, the kinetic energy of neutron at Fermi surface is 76 MeV.
Fermi energy of neutron is higher than that of a proton by $\!\approx100$MeV,
including the effect of nuclear interaction, i.e. Lane potential\cite{La62}.
From these considerations, it is seen that the electron system is in the state of a quantum degenerate with ultra-relativistic Fermi energy.
The fact that the density $\rho_{\rm e}$ is $\approx 0.004$  fm$^{-3}$, $2 \! \times \!10^{12} $ times higher
than that one in iron metal, is crucial in realizing the strong MF in the order of Tera Tesla.
It is clear by estimating the strength of MF (magnetization) constituted from aligned electrons being $\mu_{0} \mu_{\rm B} \rho_{\rm e}\approx$ 50TT with Bohr magneton $\mu_{\rm B}$.

From the above analyses, such a super many-body quantum system degenerated with ultra-relativistic Fermi energy should be treated with Dirac Hartree-Fock type calculation together with Maxwell equation.
It is not feasible, however, to handle such a gigantic system of $\approx 3 \times10^{55}$ particles.
Hence, we will solve the equation with scaled $\hbar$ method, e.g., $ \hbar_{\star}=10^{18} \hbar$ \cite{BOY95}.
This method is a generalization of Thomas-Fermi approximation in some manner, where Dirac equation
gives only number of states for quantum degenerate assemblages of $N_{\star}$ electrons.
Wave functions obtained by this equation correspond to amplitude modulation waves for standing waves and to frequency modulation waves for running waves, respectively, and describe a kind of collective mode.
We define the scaling factor $\gamma$ as $\hbar_{*}=\gamma \hbar_{\star}=\gamma 10^{18} \hbar$
(distinguish asterisk $*$ from star $\star$ in suffixes) and chose the value of $\gamma$
in consideration of size of collective motion, in which particles of $N_{\ast}$ participates.
In this paper, we take on rather large numbers of $\gamma= 1 \,\, {\rm and} \,\, 0.5$ to show
the relationship of the present method with Thomas-Fermi approximation and
to visualize the global structure of currents and MF.

The NS is rotating and has large angular momentum (AM).
The electron system shares AM with the baryon system.
We utilize the cranking model\cite{In56} to treat rotating quantum many-body system such as nucleus and Bose-Einstein condensates \cite{HOO07}.
According to the model, electrons carrying AM are located
at the peripheral region
in the vicinity of the equatorial plane and form a toroidal zone, which is very characteristic feature of rotational
motion in such a gigantic quantum degenerate many-body system.
Electrons having negative chirality in the region are selectively captured by protons through the weak interaction. Eventually, electrons having positive chirality remain and break chiral symmetry in the whole system.

Neutron density depends on the radius and is determined so as to keep local balance of $\beta$-equilibrium.
In this work, we concentrate on studying the global structure of MF and the dynamics producing the field through interaction with the electric current.
So, we assume an appropriate distribution of hadrons to determine the distribution of electrons.
For the sake of simplicity, the gravitational effects of general relativity are also ignored, since they are not important at the peripheral region.
A systematic study including hadron treated with general relativity will be presented in another paper.

In section \ref{BFC} we show the basic formulae necessary to calculate the Dirac current producing MF.
First: in subsection \ref{Basic_formulae}, we employ diagonal matrix $\beta$ called Dirac-representation.
The spinor is expressed by $j$-$j$ coupling scheme using the quantum number
$\kappa=\pm (j+\frac{1}{2})$ \cite{Ro37}.
Second:  in \ref{CHrep}, we treat  with the Weyl representation making the chiral operator $\gamma_{5}$ diagonal.
Unlike a free massless spinor, the helicity of particle in bound state is not a good quantum number, so the helicity representation\cite{JW59} is employed to describe the wave function.
In \ref{Expansion}, we expand the density and the current density
in terms of spherical harmonics and vector spherical harmonics.
The matrix elements of coupling between Dirac current and electromagnetic field are calculated with Racah algebra,
to find a simple expression in terms of $\kappa$.
We show a general form of the coupled channel equation to treat
deformation breaking spherical symmetry in \ref{cceq}.

The global structure of electromagnetic field is studied more specifically
with axially symmetric ansatz in section \ref{ElemagStrctr}.
First in subsection \ref{ElMoPo}, monopole electric polarization is considered for the quantum degenerate electron system with ultra-relativistic Fermi energy to be confined.
Quasi-bound states are introduced in \ref{Qbs} to describe the electron
in a potential deeper than $2m_{\rm e} c_{0}^{2}$ ($c_{0}$ indicates light velocity)
to circumvent the difficulty of escaping through Kleine tunnel.
We show correspondence between energies and quantum numbers of single quantum state
obtained by using different scaling factors $\gamma \!= \!1 \,\, {\rm and} \,\, 0.5$. in \ref{EngySpct}.
In \ref{ECuRNS}, the cranking model is employed to make the NS rotate together with electron system,
in which Coriolis term $-\hat{\vect{J}}\omega$ splits energies of $2j+1$ states having
different magnetic quantum number $\mu$
and causes level-crossing of two states; the one of the lowest quantum states $\mu=-j$ in the highest $j$ among
occupied states and the other of the highest $\mu=j+1$ of the lowest $j+1$ of unoccupied states.
A migration from the former to the latter 
resulting asymmetric population in time reversal partners in the vicinity of Fermi surface
breaks time reversal symmetry and brings about collective rotation. 
Convection like current is calculated for the orbiting electrons.
It is implied in \ref{crhlcyW} that an eigenstate of chirality includes  predominantly the same sign of circular helicity component. 
A numerical result of the helical current of a positive chirality eigenstate is illustrated in \ref{HlclCrrnt}.
We mention briefly the magnetic helicity in relation to the helical current in \ref{MgnHel}.
Finally, we present our scenario in \ref{ChAsHelCrr} of the origin of strong toroidal magnetic field
from two basic cause factors, i.e. the rotational motion and the chiral asymmetry. 
The sustainability of the chirality and the helicity is proven explicitly by the calculation 
of vector potential of MF produced by the spin like current . 
%
%
%
\section{Basic Formulae for Calculations}
\label{BFC}
\subsection{Dirac representation and $j$-$j$ coupling scheme}
\label{Basic_formulae}
In NS, the Fermi energy of electron is much higher than $m_{\rm e}c_{0}^{2}$. Therefore, the electrons have to be described by Dirac equation.
Wave function $\psi$ are expressed by eigen-vector of Hamiltonian $\hat{H}$
\begin{equation}
\label{Diracequatn0}
\hat{H}\psi = \left [c_{0} \vect{\alpha}  \cdot 
( \hat{\vect{p}}+ e \vect{A})-eA_{0}
+ \beta m_{\rm e} c_{0}^{2}  \right] \psi \,\, ,
\end{equation}
where $\vect{\alpha}$  and $\beta$ are $4 \times 4$  matrices and $(A_{0}(\vect{r}),  \vect{A}(\vect{r}))$ is electromagnetic potential. 
The wave function $\psi$ may be divided into two spinors $(\psi^{u} (\vect{r}\sigma),\psi^{l}(\vect{r}\sigma) )$,
upper and lower components in so called Dirac representation;
\begin{equation}
\label{Diracequatn1}
\left( \! \begin{array}{cc}
m_{0}c_{0}^{2} - e A_{0} & 
\! \displaystyle c_{0}\vect{\sigma} \! \cdot \! \left(\hat{\vect{p}}+e \vect{A} \right) 
\vspace{0.2cm} \\
\displaystyle c_{0} \vect{\sigma} \! \cdot \! \left(\hat{\vect{p}} + 
e \vect{A} \! \right) & -m_{0} c_{0}^{2} - eA_{0}
\end{array} \! \right) \!\!
\left( \!\! \begin{array}{c}
\psi^{u}(\vect{r}\sigma)  \vspace{0.2cm} \\  \psi^{\ell}(\vect{r}\sigma)
\end{array} \!\! \right) 
= E \left( \!\! \begin{array}{c}
\psi^{u}(\vect{r}\sigma) \vspace{0.2cm} \\  \psi^{\ell}(\vect{r}\sigma)
\end{array} \! \! \right)  .
\end{equation}
Using the spinors, the probability density and current density are written as
\begin{equation}
\label{dnscdns_D}
\left( \! \! \begin{array}{c} \rho(\vect{r}) \vspace{0.2cm} \\ \vect{j}(\vect{r}) \end{array} \!\! \right) \!=\!
\langle \psi \! \mid \! \!\left( \!\! \begin{array}{c}  \vect{1}_{4} 
\vspace{0.2cm} \\ c_{0} \vect{\alpha} \end{array} \! \!\right) \!\! \mid \! \psi \rangle_{\sigma}
\!\!=\!\! \left( \!\begin{array}{c} \langle \psi^{u} \mid \psi^{u} \rangle_{\sigma} \,\,
+\,\, \langle \psi^{\ell} \mid \psi^{\ell} \rangle_{\sigma}
\vspace{0.2cm} \\
\langle \psi^{u} \! \!\mid \!\! c_{0}\vect{\sigma} \!\! \mid \! \!\psi^{\ell} \rangle_{\sigma} 
\! + \! \langle \psi^{\ell} \!\! \mid \! c_{0}\vect{\sigma} \! \mid \!\! \psi^{u} \rangle_{\sigma}
\end{array} \!\!\right) . 
\end{equation}
The suffix $\sigma$ attached to the bracket $\langle \cdots \mid \cdots \rangle_{\sigma}$ expresses
calculating trace in spin matrices.
The two spinors are expanded by {\it j-j} coupling scheme and
labeleld by quantum number $\kappa$ for orbital AM $\ell_{\kappa}$
and total AM $j_{\kappa}$ defined by Rose \cite{Ro37},\cite{Ro61}  
%
%
\begin{equation}
\label{def_kappa}
\mid \kappa \mid = k = j_{\kappa}+{1 \above1pt 2} ,  \hspace{0.7cm}
\ell_{\kappa}= j_{\kappa} \pm \frac{1}{2} \hspace{0.9cm}
{\rm for} \hspace{0.3cm} 
\kappa = \pm k  .
\end{equation}
The quantum number $\kappa$ appears as eigen-value of the operator
$\hat{K} = \beta (\vect{\sigma} \!\cdot\! \hat{\vect{\ell}}+\hbar_{\ast}) $
in the four components theory \cite{He00}, where $ \hbar_{\ast} $ stands for scaled $\hbar$,
and $k$ takes on natural number. 
We introduce notations $\tilde{\ell}_{\kappa}=\ell_{-\kappa}$ ,$\tilde{\kappa}=-\kappa$ and the signature
$S_{\kappa}\equiv \kappa/ k=\ell_{\kappa}-\tilde{\ell}_{\kappa} =\pm 1$ .
The two spinors are expanded as follows
\begin{equation}
\label{Wvfnct}
\left( \begin{array}{c}
\psi^{u}(\vect{r} \sigma ) \vspace{0.2cm} \\ \psi^{\ell} (\vect{r} \sigma)
\end{array} \right)
= \left( \begin{array}{c}
\sum_{\kappa \mu} g_{\kappa \mu}(r) \chi_{\kappa \mu}(\hat{r}\sigma) \vspace{0.2cm}\\
\sum_{\kappa \mu} f_{\kappa \mu}(r) \chi_{\tilde{\kappa} \mu}(\hat{r}\sigma) 
\end{array} \right) \, .
\end{equation}
The spherical spinors $\chi_{\kappa\mu} (\hat{r} \sigma)$ are defined by
\begin{equation}
\label{sphrclspnr}
\chi_{\kappa \mu}(\hat{r} \sigma )=\sum_{\sigma=\pm \frac{1}{2}}
\langle \ell_{\kappa} m {1\above1pt 2} \sigma \mid j_{\kappa} \mu \rangle
{\cal Y}_{\ell_{\kappa}m}(\hat{r}) \chi^{( {1\above1pt 2})}_{\sigma} ,
\end{equation}
with ${\cal Y}_{\ell_{\kappa}m}(\hat{r})=i^{\ell_{\kappa}}Y_{\ell_{\kappa}m}(\hat{r})$,
differing from the Condon-Shortley convention $Y_{\ell_{\kappa}m}(\hat{r})$
by phase factor $i^{\ell_{\kappa}}$,
and $\langle\ell_{\kappa} \, m\, {1\above1pt 2} \, \sigma \mid j_{\kappa} \mu \rangle$
stands for Clebsch-Gordan coefficient.

Let us define a pseudo scalar operator
\begin{equation}
\label{sigmar}
\sigma_{r}\equiv\vect{\sigma}  \cdot \vect{n}_{r} , 
\hspace{0.7cm} {\rm with} \hspace{0.4cm} \vect{n}_{r}=(\sin \theta \cos \phi, \sin \theta \sin \phi, \cos \theta),
\end{equation}
having a property of $\sigma_{r}^{2}=\vect{1}_{2}$ and altering the parity
without changing $j$\cite{Ro61,He00},
\begin{equation}
\label{opr_sgmr}
\sigma_{r} \chi_{\kappa \mu}=-iS_{\kappa}\chi_{\tilde{\kappa} \mu} \, .
\end{equation}
We redefine radial wave function for convenience,
\begin{equation}
\left( G_{\kappa \mu}(r), F_{\kappa \mu}(r) \right) = \left( r g_{\kappa \mu}(r), r f_{\kappa \mu}(r) \right) .
\end{equation}

In case of spherically symmetric system with potential $U(r)=-eA_{0}(r)$, (necessarily $\vect{A}=0$),
the radial part of the equation turns out to be
\begin{equation}
\label{DeqDrp}
\frac{d}{dr} \left( \!\! \! \begin{array}{c}   G_{\alpha}(r) \vspace{0.3cm} \\ F_{\alpha}(r) \\
\end{array} \!\!\! \right) \! = \!
\left( \!\!\! \begin{array}{lr} 
\displaystyle \hspace{0.5cm}  -\frac{\kappa_{\alpha}}{r} & 
\hspace{-0.7cm} \displaystyle \!\!\!\!\! \frac{E \!+ \! m_{\rm e} c_{0}^{2} \! - \! U(r)}{\hbar_{*} c_{0}} 
 \vspace{0.1cm} \\
\displaystyle -\frac{E \! - \! m_{\rm e} c_{0}^{2} \! - \!U(r)}{\hbar_{*} c_{0}} &
 \hspace{-0.6cm} \displaystyle \frac{\kappa_{\alpha}}{r} \hspace{0.5cm}
\end{array} \!\! \right)
 \left( \!\! \begin{array}{c}  G_{\alpha}(r) \vspace{0.3cm} \\ F_{\alpha}(r)  \\ \end{array} \!\! \right) ,
 \end{equation}
where $\alpha$ expresses combined quantum numbers of $\kappa$ and $ \mu$.
The wave function is orthonormalized eigenfunction labeled by $\nu$ or $\nu^{\prime}$ as follows,
\begin{equation}
\sum_{\alpha} \int_{0}^{\infty} \!\!\! dr 
\left\{ G_{\alpha}^{(\nu)}(r)G_{\alpha}^{(\nu^{\prime})}(r)
+F_{\alpha}^{(\nu)}(r)F_{\alpha}^{(\nu^{\prime})}(r)
\right\}=\delta_{\nu \nu^{\prime}} \, .
\end{equation}
The above normalization in the equation using
$\hbar_{\ast}$, represents the probability of quantum assemblage
$N_{\ast}=\gamma^{3}N_{\star}$ of quantum degenerate electrons. 
%
%
%
\subsection{Weyl and Helicity Represemtations}
\label{CHrep}
To clarify the relation between chirality and helicity, we study the Weyl \cite{We29} representation of
the Dirac equation (\ref{Diracequatn1}) obtained  by the transformation
\begin{equation}
\left( \begin{array}{c} \Psi^{(\,\,\frac{1}{2})\,}(\vect{r}\sigma)  \\ \Psi^{(\!\frac{\!-\!1}{2}\!)}(\vect{r}\sigma)
\end{array} \right)=\frac{1}{\sqrt{2}}
\left( \begin{array}{c} \,\, \Psi^{u}(\vect{r}\sigma) + \Psi^{\ell}(\vect{r}\sigma)\\
  -\Psi^{u}(\vect{r}\sigma) +\Psi^{\ell}(\vect{r}\sigma)
\end{array} \right) ,
\end{equation}
where the superscript $(\pm \frac{1}{2})$  specifies chirality.
Then, Dirac equation is rewritten as
\begin{equation} 
\left( \!\!\!\! \begin{array}{cc}  
c_{0}\vect{\sigma} \! \cdot \! (\hat{\vect{p}} \! + 
\! e\vect{A} ) \! - \! eA_{0} &
\hspace{-0.8cm}  -m_{\rm e}c_{0}^{2} 
\vspace{0.1cm} \\ \hspace{-0.5cm} -m_{\rm e}c_{0}^{2} & 
\hspace{-1.0cm} - \! c_{0}\vect{\sigma} \! \cdot \! (\hat{\vect{p}} \! + \! e\vect{A} ) \! - \! eA_{0}
\end{array} \!\!\!\! \right) \!\!
\left( \!\!\!\! \begin{array}{c} \psi^{(\,\,\frac{1}{2}\,)}(\vect{r}\sigma) \vspace{0.1cm} \\
\psi^{(\! \frac{\! - \!1}{2}\!)}(\vect{r}\sigma) \end{array} \!\!\!\right) 
\! \! = \!  E \!
\left( \!\!\!\! \begin{array}{c} \psi^{(\,\,\frac{1}{2}\,)}(\vect{r}\sigma) \vspace{0.1cm} \\
\psi^{(\!\frac{\! -\!1}{2}\!)}(\vect{r}\sigma) \end{array} \!\!\!\! \right) .
\end{equation}
In case of $m_{\rm e}\!=\!0$, the four component equation is split into a decoupled pair of two-component
equations.
In the absence of an electromagnetic field, a positive chirality state is eigenvector of helicity
$\hat{h} \equiv \vect{\sigma} \cdot \hat{\vect{p}}/p$
with eigenvalue of positive helicity. On the other hand,
for the particle confined in potential $U(r)$ the helicity $\hat{h}$ is mixed.

It is shown in subsection \ref{ECuRNS} that the state relevant to forming MF has
 the highest $k$ and highest $\mu=j$. 
 Hence the plus (minus) sign of $\kappa$,
i.e., $S_{\kappa}\!=\!1$ ($\!=\!-1$), corresponds to down (up) spin.
The following operation is applied to the states orbiting a circle, 
 which rotates the spin by $\pi/2$ about  an axis $\vect{n}_{\rho} \!=\!(\cos \phi , \sin \phi, 0)$
 perpendicular to polar axis $\vect{n}_{z}$,
 \begin{equation}
 \label{prjop}
 \exp \left(-\!i\frac{\pi}{4} S_{\kappa} \sigma_{\rho} \!\right)\! =\!\cos \frac{\pi}{4} 
 - i S_{\kappa} \sigma_{\rho} \sin \frac{\pi}{4}  , 
 \hspace{1.0cm}  {\rm with} \hspace{0.2cm}  \sigma_{\rho}=\vect{\sigma}\cdot \vect{n}_{\rho} .
 \end{equation}
 Then we obtain the states considered as positive helicity state,
 in which the spin points to tangential direction of the circle.
 The negative helicity state is obtained by changing the rotation angle from $\pi/2$ to $-\pi/2$.
Since in the relevant current region, $\sin \theta$ and $\cos \theta$ are actually $\approx 1$ and $\approx 0$,
respectively, then one makes approximation like $\sigma_{\rho} \approx \sigma_{r}$,
the pseud scalar operator defined in eq.(\ref{sigmar}). 
This operation projecting out the helicity plus state from the state labelled by $\kappa$ turns out to be
%
%
\begin{equation}
\label{proj_h}
P_{\rm h}^{(\!\pm\!)}\chi_{\kappa \mu}(\hat{r},\sigma) 
\!=\!\frac{1}{\sqrt{2}} \left( 1\mp iS_{\kappa}\sigma_{r} \right) \chi_{\kappa \mu}(\hat{r},\sigma) 
 \!=\! \frac{1}{\sqrt{2}} \left( \chi_{\kappa \mu}(\hat{r},\sigma) \mp \chi_{\tilde{\kappa} \mu}(\hat{r},\sigma) \right) .
\end{equation}

From eq.(\ref{opr_sgmr}), this equation gives a unitary transformation between complementary states of
helicity representation, introduced for the analysis of collisions of relativistic particles with spin \cite{JW59},
and $j$-$j$ coupling states (canonical representation).
We chose negative value of $\kappa$ in eq. (\ref{proj_h}) to fit the phase convention\cite{BM69},  e.g. ,
\begin{equation}
\label{cir_hel}
\chi_{hjm}(\hat{r},\sigma) \! \equiv \! \frac{1}{\sqrt{2}} \!
\left( \! \chi_{\!-\! k \mu}(\hat{r},\sigma)  -  (\!-1\!)^{h-\frac{1}{2}}\chi_{k \mu}(\hat{r},\sigma)\right) ,
\end{equation}
where $h=1/2$ is taken on for positive helicity and $h=-1/2$ for negative helicity.
To distinguish the helicity $\hat{h}$
from that of helicity representation in eq.(\ref{cir_hel}),
we will call the former as linear helicity and the latter
as circular helicity.
To solve each equation, the spinors of helicity representation are employed,
\begin{equation}
\label{WylWvf}
\psi^{(c)}(\vect{r} \sigma)  = \sum_{h=\pm \frac{1}{2}} \sum_{j\,\mu}
u^{(c)}_{hj\mu}(r)\chi_{hj\mu}(\hat{r} \sigma)  .
\end{equation} 

Now, we redefine $H^{(c)}_{hj\mu}(r) =  r u^{(c)}_{hj\mu}(r) $ and operate $\vect{\sigma}\cdot \hat{\vect{p}}$ to
$\psi^{(c)}(\vect{r} \sigma) $ to obtain
\begin{equation}
\label{sgmp}
\vect{\sigma} \! \cdot \! \hat{\vect{p}} \left(  \frac{H^{(c)}_{hj\mu}(r)}{r} \right)\chi_{hj\mu}
 =   (\!-\!1)^{h-c}
\frac{\hbar_{\ast}}{r} \! \left(\frac{dH^{(c)}_{hj\mu}}{dr} \chi_{-hj\mu}
+\frac{k}{r} H^{(c)}_{hj\mu}(r) \chi_{hj\mu} \! \right) .
\end{equation}
From the calculation, it is seen that only radial motion mixes the circular helicity.
 
The probability density and current density are expressed as
\begin{equation}\label{dncrdn_W}
\left( \begin{array}{c} \rho(\vect{r}) \vspace{0.2cm} \\ \vect{j}(\vect{r}) \end{array} \right) = 
\left( \begin{array}{c} \rho^{(\!+\!)}(\vect{r}) + \rho^{(\!-\!)}(\vect{r}) \vspace{0.1cm} \\ 
\vect{j}^{(\!+\!)}(\vect{r}) +  \vect{j}^{(\!-\!)}(\vect{r}) \end{array}  \right)  ,
\end{equation}
with
\begin{equation}\label{chrlcrrnt}
\left( \begin{array}{c} \rho^{(\!\pm\!)}(\vect{r}) \vspace{0.1cm} \\ 
\vect{j}^{(\!\pm\!)}(\vect{r}) \end{array} \right) =
\left( \begin{array}{c} \langle \psi^{(\!\pm)\!} (\vect{r},\sigma)\mid 
\psi^{(\!\pm\!)}(\vect{r},\sigma)\rangle_{\sigma} 
\vspace{0.1cm} \\ 
\langle \psi^{(\!\pm\!)} (\vect{r},\sigma)\! \mid \pm c_{0} \vect{\sigma} \! 
\mid \! \psi^{(\!\pm\!)} (\vect{r},\sigma) \rangle_{\sigma}
\end{array} \right) . 
\end{equation}
In the Weyl representation, the matrix $\vect{\alpha}^{\prime}$
is diagonal in chirality, 
and therefore the current density is a separable sum of each component,
rather than transitions between upper and lower components
in the Dirac representation.
%
%
\subsection{Moments of Density and Current Density}
\label{Expansion}
The density and current density are expanded like
\begin{equation}
\left\{ \begin{array}{rl}
\rho(\vect{r}) & =\sum_{LM}\rho_{LM}(r) {\cal Y}_{LM}(\hat{r}) \vspace{0.2cm} \\
\vect{j}(\vect{r}) & =\sum_{LJM}c_{0}\eta_{LJM}(r) \vect{\cal V}_{LJM}(\hat{r}) 
\end{array} \right. ,
\end{equation}
in terms of spherical harmonics and vector spherical harmonics as follows,
where $\vect{\cal V}_{LJM} (\hat{r})$ is defined as \cite{He00,BW52}
\begin{equation}
\vect{\cal V}_{LJM} (\hat{r})=\sum_{M_{L} S}\langle LM_{L} 1 S \mid J M\rangle
{\cal Y}_{LM_{L}}  (\hat{r}) \vect{\chi}_{S}^{(1)}  .
\end{equation}
The notation $\vect{\chi}_{S}^{(1)} $ is tensor representation of rank 1. 
For the same $J$, there are three functions with different $L (=J \pm 1 ,J)$. 
In time independent states, the current density is divergent free, i.e.,
$\nabla \cdot \vect{j}=0$.
In case of $L=J$, a vector field
\begin{equation}
\vect{\cal C}_{JM}(\vect{r},{\cal M})=c_{J}(r)\vect{{\cal V}}_{JJM}(\hat{r}) \, ,
\end{equation}
is shown to be divergent free and  to have parity $(-1)^{J+1}$.
It may be called the magnetic $2^{J}$-pole field
in analogy with the multipole radiation field \cite{He00} \cite{BW52}.
A linear combination of two others, $(L=J\pm 1)$ being solenoidal, is defined
\begin{equation}
\vect{{\cal C}}_{JM}(\vect{r},{\cal E})\!=\! C_{J}^{(\!-\!)} k_{J\!-\!1}(r)
\vect{{\cal V}}_{(\!J\!-\!1)\!\, J \!M}(\hat{r})\!+\!
C_{J}^{( \! + \! )}k_{J\!+\!1}(r)\vect{{\cal V}}_{\!(\!J\!+\!1\!)\!\, J\! M}(\hat{r}) \,  ,
\end{equation}
with
\begin{equation}
( C_{J}^{(\!+\!)} \, , \,C_{J}^{( \! - \! )} )\equiv
\left( \sqrt{\!\frac{J}{2J\!+\!1}\!} \, , \, \sqrt{\!\frac{J\!+\!1}{2J\!+\!1}\!}  \, \right) \, .
\end{equation}

By using the orthonormal properties of spherical harmonics and vector spherical harmonics,
the moments of density and current density are calculated as
\begin{equation}
\left( \begin{array}{c}
\rho_{JM}(r)  \vspace{0.2cm} \\
\eta_{LJM}(r)
\end{array} \right)=\sum_{\alpha \beta}
\left(\begin{array}{c} \hspace{0.7cm}
C_{\alpha \beta}^{LJ}
 \rho_{\alpha \beta}(r) \vspace{0.2cm} \\
\xi_{\kappa_{\alpha} \tilde{\kappa}_{\beta}}^{LJ}
C_{\alpha \tilde{\beta}}^{LJ}
\eta_{\alpha \beta}(r)
\end{array} \right) \, ,
\end{equation}
together with
\begin{equation}
\label{dfdnstymtx}
\left(  \!\! \begin{array}{c} 
\rho_{\alpha \beta}(r) \vspace{0.2cm} \\
\eta_{\alpha \beta}(r) 
\end{array} \!\! \right) \! = \!\!\frac{1}{r^{2}}
\sum_{ \nu (\in{\rm occ})}
\left( \begin{array}{c}
G_{\alpha}^{(\nu)}(r)G_{\beta}^{(\nu)}(r)+
F_{\alpha}^{(\nu)}(r)F_{\beta}^{(\nu)}(r)
\vspace{0.2cm} \\
G_{\alpha}^{(\nu)}(r)F_{\beta}^{(\nu)}(r)+
F_{\alpha}^{(\nu)}(r)G_{\beta}^{(\nu)}(r)
\end{array} \right) \, ,
\end{equation}
where $\nu$ specifies occupied quantum states.
The kinematical factor $C_{\alpha \beta}^{LJ}$ is given by
\begin{equation}
C_{\alpha \beta}^{LJ}=
D_{\kappa_{\alpha} \kappa_{\beta}}^{LJ}(-1)^{j_{\beta}-\mu_{\beta}} \!
\langle j_{\alpha} \mu_{\alpha} j_{\beta}\!-\! \mu_{\beta}  \!\! \mid \! J M\rangle  \, ,
\end{equation}
and 
$D_{\kappa \kappa^{\prime}}^{LJ}$ represents
\begin{equation}
\label{rdmtrx}
D_{\kappa \kappa^{\prime}}^{LJ} \equiv \displaystyle
\frac{\langle \ell_{\kappa} \frac{1}{2}  j_{\kappa} \parallel  {\cal Y}_{L} \parallel
 \ell_{\kappa^{\prime}} \frac{1}{2} j_{\kappa^{\prime}} \rangle }{\sqrt{2J+1}}=
S_{\kappa}S_{\kappa^{\prime}}
\cos{\left( \frac{\pi}{2}(\ell_{\kappa^{\prime}} \!\!- \!\! L
\! - \! \ell_{\kappa})\right)} d^{J}_{k \,k^{\prime}} ,
\end{equation}
and
\begin{equation}
d^{J}_{k \, k^{\prime}}=
\sqrt{\! \frac{(2 j_{k} \! + \!1)(2 j_{k^{\prime}} \! + \! 1)}{4\pi(2J+1)}}
( \! -1 \! )^{j_{k^{\prime}} - \frac{1}{2} }
\langle j_{k} {1\above1pt 2} \,  j_{k^{\prime}} - \! {1 \above1pt 2} \mid \! J0 \rangle ,
\end{equation}
with
\begin{equation}
\label{def_xi} 
 \xi_{\kappa \kappa^{\prime}}^{LJ} \!=\! -\hat{J}^{-1}\!\!
\left\{  \!\! \begin{array}{l} 
\displaystyle C_{J}^{(\mp)\!} (\kappa +\kappa^{\prime}) \mp C_{J}^{( \! \pm \! )}\hat{J}
\vspace{0.2cm}\\ \displaystyle 
\kappa -\kappa^{\prime}  \hspace{1.2cm}
\end{array}
\right.
 {\rm for} \hspace{0.1cm} \left\{ \!\!
\begin{array}{l} J \! = \! L\pm1  \vspace{0.1cm} \\ J=L 
\end{array}  \right. \, ,{\rm and} \hspace{0.1cm} \hat{J} \! = \!\sqrt{\!J(J\! + \! 1)}
\end{equation}
 is independent of magnetic quantum numbers
$\mu_{\alpha}, \mu_{\beta} \,\, {\rm and } \,M$ due to Wigner-Eckart theorem,
where
$\langle \ell_{\kappa} \frac{1}{2}  j_{\kappa} \! \parallel \! {\cal Y}_{L} \!\! \parallel
 \ell_{\kappa^{\prime}} \frac{1}{2} j_{\kappa^{\prime}} \rangle$
is called as reduced matrix element.
%
%
%
\subsection{Coupling Matrices of Electromagnetic Field}
\label{cceq}
Next, let us expand electromagnetic potentials $A_{0} (\vect{r})$ and $\vect{A}(\vect{r})$ 
in terms of spherical and vector spherical harmonics as follows,
\begin{equation}
\left\{ \begin{array}{l}
A_{0}(\vect{r}) = \sum_{LM}R_{LM}(r) {\cal Y}_{LM}(\hat{r}) \vspace{0.2cm} \\
\vect{A}(\vect{r})=\sum_{LJM} H_{LJM}(r)\vect{\cal V}_{LJM}(\hat{r})
\end{array}
\right. .
\end{equation}
From Maxwell equation, the potentials are given by the densities through the Green's function of
Poisson equation as
%
%
\begin{equation}
\label{Green_fn}
\left(  \begin{array}{c} R_{LM}(r) 
\vspace{0.2cm} \\ H_{LJM}(r) \end{array}  \right) 
\! =\frac{-eN_{\ast}}{2L+1} 
\! \int_{0}^{\infty} \!\! r^{\prime\, 2}dr^{\prime} \frac{(r_{<})^{L}}{r_{>}^{L+1}} \!\!
\left( \!\!\! \begin{array}{c} \! \varepsilon_{0}^{-1}\rho_{LM}(r^{\prime})
\vspace{0.2cm}  \\  \mu_{0} c_{0}\,\eta_{LJM}(r^{\prime}) \end{array} \!\!\!  \right) \, ,
\end{equation}
with
\begin{equation}
\label{argGrf}
(r_{<},r_{>})=\left\{ (r^{\prime} , r)\,\, {\rm or} \,\, (r, \, r^{\prime}) \right\}
 \hspace{0.3cm} {\rm for} \hspace{0.1cm} 
 \left\{  r^{\prime} < r \,\, {\rm or} \,\,  r < r^{\prime}  \right\} ,
\end{equation}
where $\varepsilon_{0}$ and $\mu_{0}$ express dielectric constant and magnetic permeability of the vacuum,
respectively. 
It is noted again that $H_{LJM}(r)$ is real 
and depends on only $\eta_{LJM}(r)$, because the Laplacian included in Poisson equation is a scalar operator.
 
Finally we obtain Dirac equation in the coupled channel form,
\begin{equation}
\label{cpcheq}
\begin{array}{l}
\left( \begin{array}{l}
\displaystyle
\frac{dG_{\alpha}(r)}{dr}   + \frac{\kappa_{\alpha}}{r}G_{\alpha}(r) 
-\sum_{\beta} V_{\tilde{\alpha} \beta}(r) G_{\beta}(r)  \\ \displaystyle
\frac{dF_{\alpha}(r)}{dr}   - \frac{\kappa_{\alpha}}{r}F_{\alpha}(r)
-\sum_{\beta} V_{\alpha \tilde{\beta}}(r) F_{\beta}(r)  
\end{array}  \right)  \vspace{0.1cm} \\   \displaystyle \hspace{1.1cm}
=\frac{1}{\hbar_{\ast} c_{0}} \left(
\begin{array}{l}
\displaystyle
   \,\,\left( E + m_{\rm e} c_{0}^{2} \right) F_{\alpha}(r) + 
 \sum_{\beta}U_{\tilde{\alpha} \tilde{\beta}}(r) F_{\beta}(r)  , \\
\displaystyle \!\!\!
  - \left( E - m_{\rm e} c_{0}^{2} \right)G_{\alpha}(r)  - \sum_{\beta}U_{\alpha \beta}(r) G_{\beta}(r) .
\end{array}  \right) \, .
\end{array}
\end{equation}   
The suffixes $\tilde{\alpha}$ indicate the dual state of $\alpha$, i.e., 
$(\ell_{\tilde{\alpha}}=\tilde{\ell}_{\alpha})$.
The matrix elements are given by
\begin{equation}\label{CpMtx}
\left( \begin{array}{c} U_{\alpha \beta}(r) 
\vspace{0.2cm} \\ V_{\alpha \beta}(r) \end{array} \right) 
= \frac{e}{\hbar_{\ast}}
\sum_{LJM} \! C_{\alpha \beta}^{LJ} 
\left( \begin{array}{c}
R_{LM}(r)  \vspace{0.2cm} \\ 
\xi_{\kappa_{\alpha} {\kappa}_{\beta}}^{LJ} 
H_{LJM}(r) 
\end{array}  \right) \, .
\end{equation}
Mixing channel of different AM breaks spherical symmetry to describe deformed states.
%

%
\section{Electromagnetic Structure of NS}
\label{ElemagStrctr}
\setcounter{equation}{0}
\subsection{Electric Monopole Polarization}
\label{ElMoPo}
The electron system in NS is quantum degenerate with ultra-relativistic
Fermi energy is confined in the star.
The atomic nucleus fails to hold electrons inside of the body. 
While the size of nucleus, e.g. 6 fm, is much smaller than Bohr radius,
the size of NS seems large enough to accommodate electrons.
But still some force is necessary to confine electrons. 
It is only Coulomb force, which is passed over behind the local charge neutrality anzats.

In practice the density of neutron is strongly dependent on radius together with those of proton and electron.
Assuming Thomas-Fermi approximation,
the potential $U(r)$ is determined  by
Fermi energy depending on local density
$\rho_{\rm e}(r)$ through \vspace{-0.2cm}
$$U(r)=\sqrt{ (\hbar c_{0} (3 \pi^{2}\rho_{\rm e}(r))^{1/3})^{2} + m_{\rm e}^{2} c_{0}^{4}} .
\vspace{-0.2cm} $$
Therefore proton density is mostly same as electron's due to local charge neutralization.
A slightly deferent charge distribution of proton from electron gives rise to
monopole charge polarization $\rho^{({\rm pol})}(r)$, which is the source term of Poisson equation
\begin{equation}
  \frac{\varepsilon_{0}}{e} \frac{1}{r^{2}} \frac{d}{dr}r^{2}\frac{d}{dr}U(r)= -\rho^{({\rm pol})}(r) .
\vspace{-0.1cm}  
\end{equation} 

For the convenience of calculation, let us use the following type of potential,   
\vspace{-0.2cm}
\begin{equation}
\label{cntlpt}
U(r)=U_{0}\left\{ 1+\left(\frac{r}{R_{\star}}\right)^{\!\! \lambda} \right\}^{\!\! -1} \, .
\vspace{-0.3cm}
\end{equation}
If $r$ is much larger than radius $R_{\star}$, the potential decreases as $r^{-\lambda}$.
For large $\lambda$, $U(r)$ is flat in the region of $r<R_{\star}$ and decreases sharply in $r>R_{\star}$.
The radial component of electric field $E_{r}(r)$ is expressed as
\vspace{-0.2cm}
\begin{equation}
E_{r}(r)=-\frac{U_{0}\lambda}{er} \left( \frac{r}{R_{\star}} \right)^{\!\! \lambda}
\!\! \left\{1+\left(\frac{r}{R_{\star}}\right)^{\!\! \lambda} \right\}^{\!\!-2} \, ,
\vspace{-0.2cm}
\end{equation}
the polarization charge density turns out to be
\begin{equation}
\rho^{({\rm pol})}(r)  =  \frac{\varepsilon_{0}}{r}E_{r}(r)
\left\{ \! (  \lambda \! + \!1  ) \! - \! (  \lambda \! - \! 1 )
\left( \! \frac{r}{R_{\star}} \!\right)^{\! \lambda}  \right\}
\left(\! 1 \! + \! \left( \! \frac{r}{R_{\star}} \! \right)^{\! \lambda} \right)^{\!-1} \, .
\end{equation}
The electric field is the strongest at $r_{\rm \!mx}=\sqrt[\lambda]{(\lambda-1)/(\lambda+1)}R_{\star}$.
In case of $\lambda =4$ ($=6$), $r_{\rm \! mx}=0.88 R_{\star}$ ( $=0.95 R_{\star}$).
The electric field and charge density at $r_{\rm \! mx}$ are given by
\begin{equation}
E_{r}(r_{\rm \! mx}) \! = \! \frac{U_{0} }{e r_{\rm \! mx}}\frac{\lambda^{2} \! - \! 1}{4\lambda} ,
\hspace{0.5cm} {\rm and} \hspace{0.5cm}
\rho^{({\rm pol})}(r_{\rm \! mx}) \! = \! \frac{2\varepsilon_{0}}{r_{\rm \! mx}}E_{r}(r_{\rm \! mx})  ,\hspace{-0.2cm}
\end{equation}
respectively.
Let us take on the values of $U_{0}$, $R_{\star} $ and $\lambda$ as -160MeV, 12km and 4, respectively.
Then,we obtain $r_{\rm mx}=10.6$km,
$E_{r}^{\rm (max)}\!\!=1.4 \!\times \!10^{4}$Vm$^{-1}$,
$\rho^{({\rm pol})}(r_{\rm mx})=2.4 \times 10^{-11}\!$ \!Cm$^{-3}$, corresponding to the electron number
$1.5 \times 10^{8} {\rm m}^{-3}=1.5 \times 10^{-37} $fm$^{-3}$.
This number is negligibly small compared to $\rho_{\rm e}(r_{\rm mx})=0.0044$fm$^{-3}$, namely
$\rho^{\rm (pol)}(r_{\rm mx})/e\rho_{\rm e}(r_{\rm mx})=3.4 \times 10^{-35}$.

This small number is found in gravitational force compared to electrostatic force between
two protons,  i.e., 
$Gm_{\rm p}^2 / \left(e^{2}/4\pi \varepsilon_{0}\right)=8.093 \times 10^{-35}$.
So it is interesting to consider the Newtonian gravitational  force and electrostatic force
acting on a proton at $r_{\rm mx}$. 
The former and the latter are estimated as 2.4\,f N and 2.3\,f N, respectively,
assuming mass inside of sphere of radius $r_{\rm mx}$ to be 80 \%.
This similarity in the strength seems more than a coincidence.

According to above considerations, it is seen that only a tiny fraction, $10^{-35}$, of electrons contributes
to forming electrostatic field to confine the major part of the other high energy electrons.
It would be thought the local neutrality satisfied.
At the same time, the electric field induced by this polarization
affects proton pushed out against the gravitational force.
The proton density pushed outward produces a force pushing out the neutrons through the
gradient of Lane potential\cite{La62}(isovector potential) resulting in a reduction of pressure inside of NS.
The reduction of pressure may be important in the discussions on the equation of state for nuclear matter,
which is intensively argued in the context of stability in the massive NS such as $2M_{\odot}$.
It is necessary to study more systematically including general relativity together with
electromagnetic force as well as hadronic matter.
%
\subsection{Quasi Bound States}
\label{Qbs}
\begin{figure}
\centerline{\includegraphics[width=12 cm,height=5.2 cm]{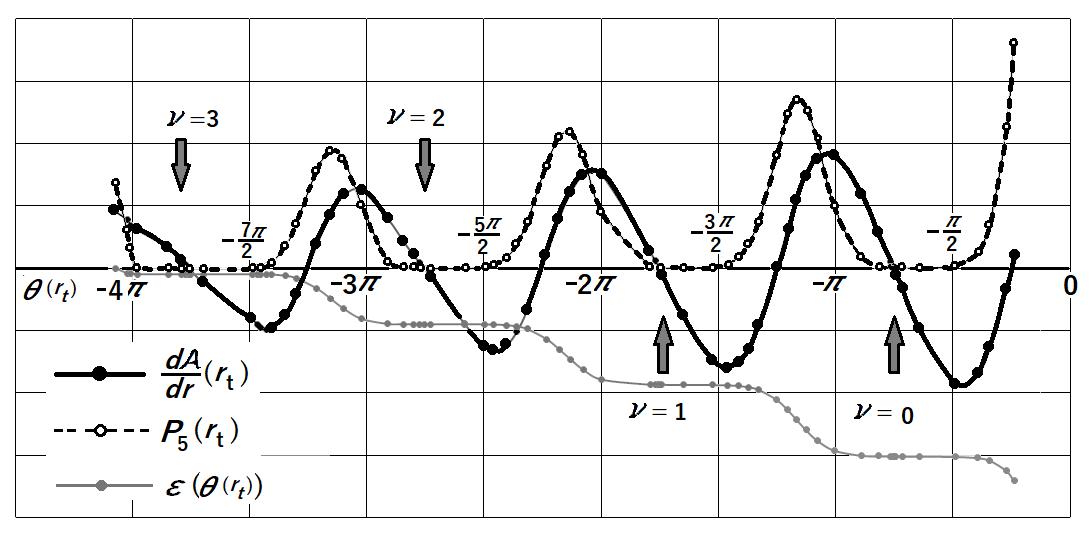}}
\caption{Probability ratio of the boundary point to that of inside of the point, 
$P_{\kappa}(r_{\rm t})$ is plotted as a function of $\theta_{\kappa}(t_{\rm t})$.
arrows are pointing the $\theta(r_{\rm t})$ at $d{\cal A}/dr=0$ for each radial node
number $\nu$.}
\label{bndrycn}
\end{figure}

The electron bounded in the potential deeper than $2m_{\rm e} c_{0}^{2}$ has always a probability to escape through Klein tunnel, similarly to alpha decay in nuclei heavier than mass number $A \approx 150$, where
alpha particle decays through, let's say, Gamow tunnel.
To study the behavior of wave function in the Klein tunnel, we deal with Dirac equation (\ref{DeqDrp})
together with the central potential in eq. (\ref{cntlpt}) .
The escaping probability may be reduced by tuning the energy $E$ to make the amplitude at a certain point
$ r_{\rm t}$ small compared to the probability inside of it, that is
\begin{equation}
P_{\kappa}(r_{t}) = \frac{r_{t} {\cal A}_{\kappa}^{2}(r_{t})}{\int_{0}^{r_{t}}{\cal A}_{\kappa}^{2}(r) dr} .
\end{equation}
\begin{figure}
\centerline{\includegraphics[width=14 cm,height=5.2 cm]{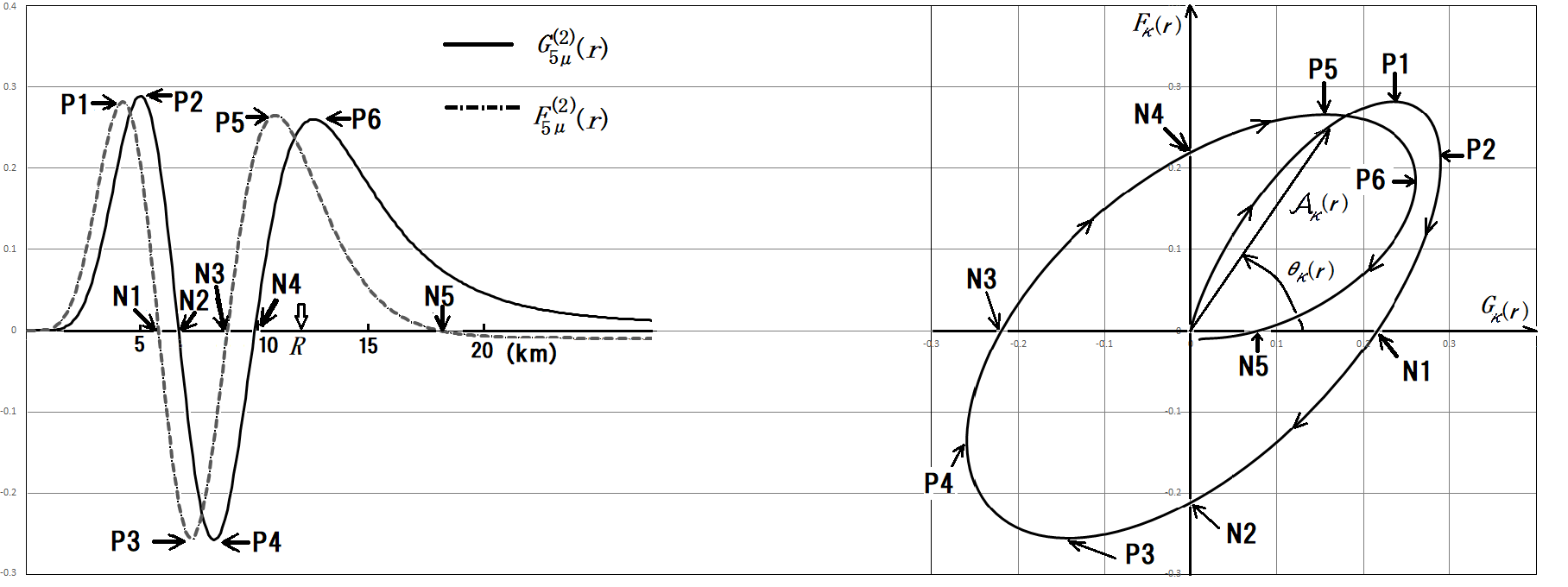}}
\caption{Wave functions of upper and lower component for $\kappa=5, \nu=2$ are depicted in 
two different ways in l.h.s. and their AAr. is illustrated in r.h.s.}
\label{wvfc52}
\end{figure}
Here we describe a set of amplitudes of upper and lower components shown in the left hand side (l.h.s.)
of Fig \ref{wvfc52} as  
\begin{equation}
\left( \begin{array}{c} G_{\kappa}(r) \vspace{0.1cm} \\ F_{\kappa}(r) \end{array}\right)
= {\cal A}_{\kappa}(r) 
\left( \begin{array}{c} \cos \theta_{\kappa}(r)\vspace{0.1cm} \\ 
\sin \theta_{\kappa}(r) \end{array} \right), \hspace{0.4cm} {\rm with}
\hspace{0.3cm} {\cal A}_{\kappa}(r) = \sqrt{G_{\kappa}(r)^{2}+F_{\kappa}(r)^{2}} ,
\end{equation}
illustrated in the right-hand side (r.h.s.) of Fig. \ref{wvfc52}, for $\kappa=5$ and $\nu=2$.
We call this representation as Angle-Amplitude representation, abbreviated as AAr.
The orbit starts from the origin and, the radious ${\cal A}(r)$ increases until the point
between two points indicated by P1 and P2. 
From the Dirac equation (\ref{DeqDrp}), we obtain
\begin{equation}
\frac{1}{{\cal A}_{\kappa}(r)}\frac{d{\cal A}_{\kappa}(r)}{dr}
=-\frac{\kappa}{r} \cos 2 \theta_{\kappa}(r)+m\sin 2\theta_{\kappa}(r) ,
\end{equation} 
and
\begin{equation}
\frac{d \theta_{\kappa}(r)}{dr}=-(\varepsilon -u(r))+
\frac{\kappa}{r} \sin 2 \theta_{\kappa}(r)+m\cos 2\theta_{\kappa}(r) ,
\end{equation}
with $\varepsilon=E/\hbar_{*}c_{0}$, $u(r)=U(r)/\hbar_{*}c_{0}$
and $m=m_{\rm e}c_{0}/\hbar_{*}$ respectively.
The angle variable $\theta_{\kappa}(r)$, as a function of $r$ for a given $\varepsilon$, 
starts from $\pi/2$ at the origin $r=0$, and 
is decreasing, i.e., going around clockwise, until the radial coordinate $r$ gets closer to the point
$r_{\rm c}$ satisfying $\varepsilon = u(r_{\rm c})$.
Then we set up the end point $r_{t}$ as $ d\theta/dr |_{r=r_{\rm t}} \!\!\!=0 $.
The angle $\theta_{\kappa}(r_{\rm t})$ is monotonously decrease function of $\varepsilon$.
The probability ratio $P_{\kappa}(r_{\rm t})$ oscillates as a function of $\theta(r_{\rm t})$ 
and reaches zero line 
when $d{\cal A}/dr $ crosses the zero line with falling right shoulder
as shown in Fig. \ref{bndrycn}  with thick arrows.
We determine the energy of quasi bound states by tuning $\varepsilon$ to make
$d{\cal A}/dr|_{r=r_{\rm t}} \!\!\! =0$ hold.
%

The radial node number $\nu$ is determined by counting points
where the upper component $G_{\kappa}^{(\nu)}(r)$ becomes zero, the points indicated as
N2 and N4 in Fig.\ref{wvfc52}.
%
%
\subsection{Energy Level Density with different Scaling Factors}
\label{EngySpct}
\begin{figure}
\centerline{\includegraphics[width=14 cm,height=4.5 cm]{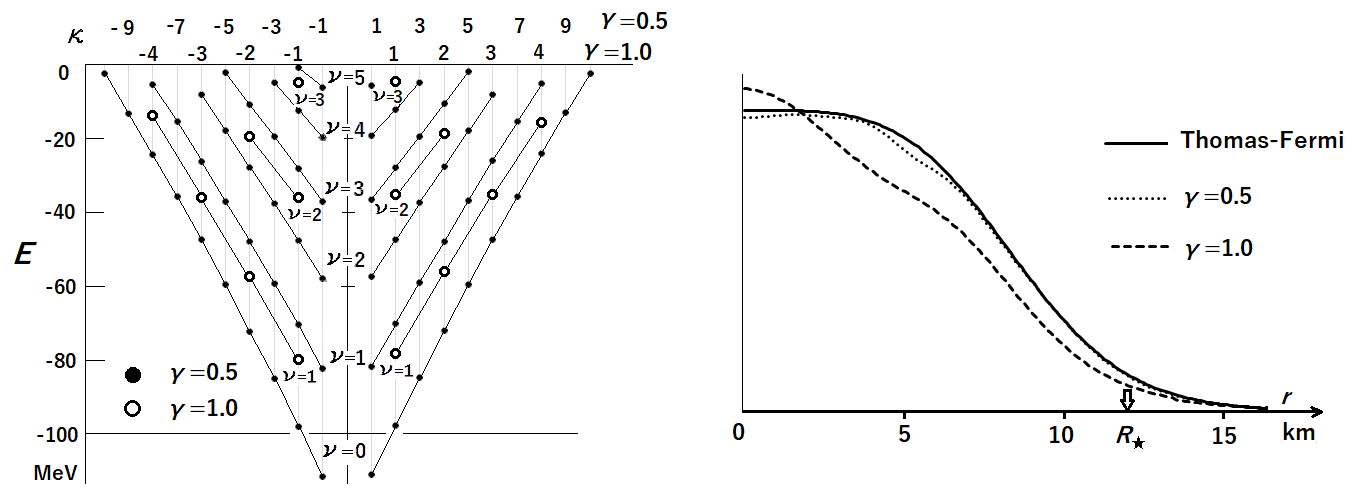}}
\caption{Energy spectra versus $\kappa$ are plotted in l.h.s. for different
scaling factors $\gamma=1$ and $\gamma=0.5$. The comparison of density radial distributions
of two scaling factors to that of Thomas-Fermi approximation are shown in r.h.s..}
\label{sclng}
\end{figure}

In order to examine the validity of the scaled $\hbar$ method, we solved Dirac equation
with taking  on different values of scaling factor $\gamma$ and obtained the quasi-bound states.
Their energy spectra are plotted as a function of $\kappa$ for the scaling factors
$\gamma=1$ and $\gamma=0.5$, in case of $U(0)=-160 {\rm MeV}$, in l.h.s. Fig. \ref{sclng}.
The points of energy spectra are connected by lines for the same radial node number $\nu$,
separately for different two scaling factors. It is found in l.h.s. of
Fig. \ref{sclng}, that the lines are almost parallel,
indicating that the scaled $\hbar$ method works well.
In other words, angularly node number, i.e., AM
and radial node number are proportional.
As for the density distributions for different scaling factor, the smaller $\gamma$ is more accurate than
the large number, and the density profiles converge closer to Thomas-Fermi distribution for smaller $\gamma$,
shown in r.h.s. of Fig. \ref{sclng}.
%
%
%
\subsection{Convection like Current in Rotating NS}
\label{ECuRNS}
The observed NS is rotating and therefore has the total AM.
It is slightly decreasing in time due to magnetic dipole radiation,
but within a characteristic period of motion of materials in NS, 
it is conserved and is shared by the electron system.
In the limit of weak coupling between electrons and hadrons,
the total energy and the AM of hadrons and electron are
expressed as
$E_{\rm tot}=E_{\rm h}+E_{\rm e}$ and
$I_{\rm tot}=I_{\rm h}+I_{\rm e}$, respectively. 
Therefore to make the total energy minimum, 
angular velocity is considered as the Lagrange multiplier
for the total AM conservation.
The partial AM is determined so as to minimize the total energy like,.e.g.,
$
\omega = \partial E_{\rm h} /\partial I_{\rm h}
=  \partial E_{\rm e} /\partial I_{\rm e}.
$

For simplicity, the electron system is axially symmetry along the rotating axis.
The effects of rotation with angular velocity $\omega$ are taken into account by Coriolis force
in the cranking model utilized in quantum degenerate many-body systems such as
nuclei \cite{In56} and Bose-Einstein condensates \cite{HOO07}\hspace{0.1cm},
where the single quantum state energy $E$ is just replaced by $E-\hbar_{\ast}\omega \mu_{\alpha}$.
 
This Coriolis term, being time reversal odd, 
splits $2j_{\alpha}+1$ degenerate states in the $j_{\alpha}$ level, namely higher $\mu$ (positive) levels
are pushed down and lower $\mu$ (negative) levels are pushed up.
Those levels have equal spacing with $\hbar_{*}\omega$ like Zeeman splitting.
As angular velocity is increased, the lowest $\mu(=-j_{\alpha})$ level of occupied highest $j_{\alpha}$ state
just below Fermi energy crosses with the highest $\mu^{\prime}(=j_{\alpha}+1)$ level
of unoccupied $j_{\alpha}+1$ state just above Fermi energy.
At the crossing point, the transition from the occupied state to the unoccupied state causes
particle-hole like excitation corresponding to a collective rotational motion.
This rearrangement in occupation breaks spherical and time reversal symmetries of the entire system, and
gives rise to circular current density in peripheral due to the highest $j_{\alpha}$
and near the equatorial plane due to the highest $\mu=j_{\alpha}$.

When both of time reversal partners are occupied, these quantum states do not contribute to
AM and current. 
Hence we may consider the above particle-hole like excitation to be the rotational motion of
electron and estimate the current flux with wave function of the single quantum state
by choosing an appropriate scaling factor $\gamma$. 
The number $N_{*}$ represents the number of electrons participating the collective rotation. 
We calculate energies of the highest $k$ occupid state, $E_{k}$ and of the lowest $k+1$ unoccupied
state, $E_{k;1}$  for five points of different values of $\gamma$ (see Table \ref{trdlfld}) 
and find the difference $ \Delta E=E_{k+1}- E_{k} \approx 21.6 \gamma $MeV, 
being proportional to $\gamma$. 
Therefore, the rotation period $P$ is related with $\gamma$ by $P=2.1 /\gamma$ msec.   

From the dependence of $\kappa$ and $\kappa^{\prime}$  in the factor of reduced matrix elements,
$\xi^{L\,J}_{\kappa \, \kappa^{\prime}}$ in eq. (\ref{def_xi}),
it is seen that only magnetic odd multipolarity having even parity, i.e., $J=L$, contributes to the current.
In a axially symmetric system $\mu$ is eigenvalue and therefore $M=0$, so that the vector spherical harmonic
under the condition has a following property,
\begin{equation} 
\vect{\cal V}_{LL0}(\hat{r})=-i{\cal Y}_{L1}(\theta 0) \vect{n}_{\phi} \, ,
\hspace{1.0cm} {\rm with } \hspace{0.3cm} \vect{n}_{\phi}= (-\sin \phi , \cos \phi, 0) .
\end{equation}
\begin{figure}
\centerline{\includegraphics[width=11 cm,height=4 cm]{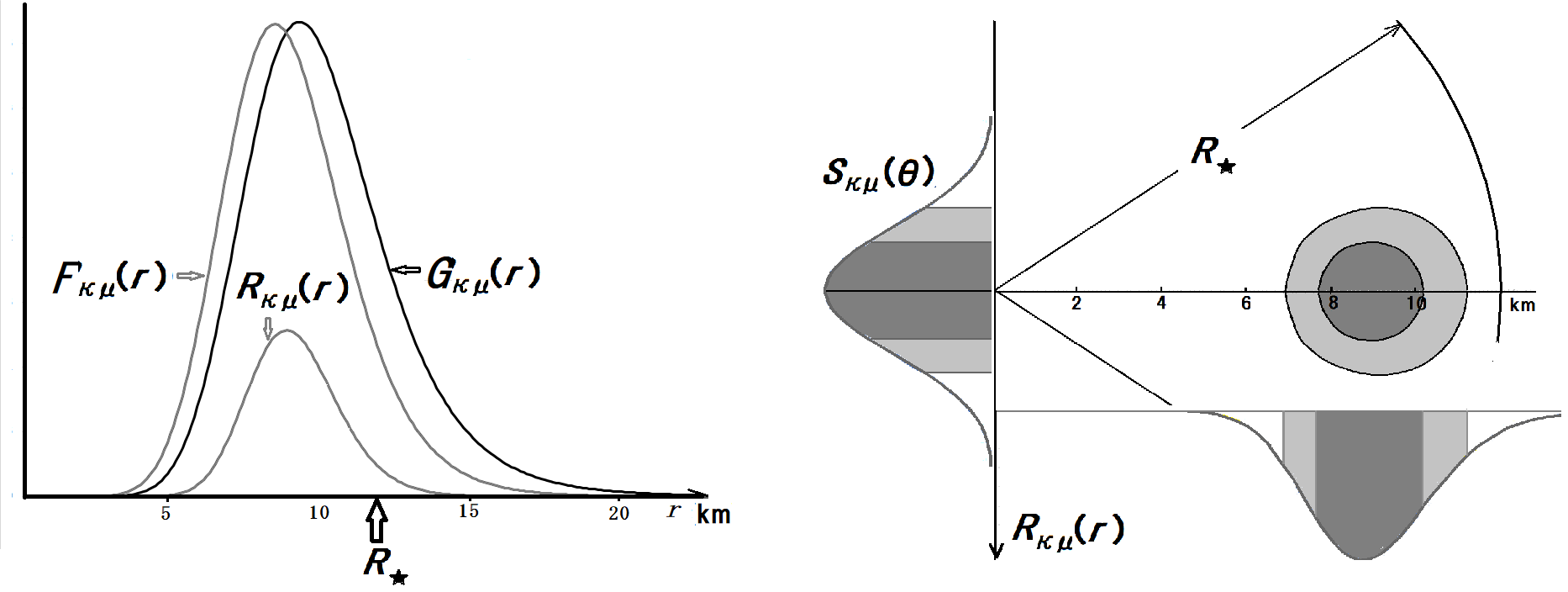}}
\caption{Radial wave functions $G_{\kappa \mu}(r)$, $F_{\kappa \mu}(r)$ and
transition amplitude $R_{\kappa \mu}(r)$ for $\kappa=10$ in l.h.s.
The cross section
of convection like current on the meridian plane is shown in r.h.s.}
\label{crntfld}
\end{figure}
The probability current of the orbital $(\kappa\!=\!j\!+\!\frac{1}{2} \,\, \mu\!=\! j)$
has only component of tangent
of circles and makes toroidal zone having density of cross section $q_{\kappa \mu}(r \theta)$
\begin{equation}
\vect{j}_{\kappa \, \mu}(\vect{r})
=q_{\kappa \mu}(r \theta) \vect{n}_{\phi} \,  .
\end{equation}
It is a function of $ r$ and $\theta$ having the following separable form
\begin{equation}
q_{\kappa \mu}(r \theta) = R_{\kappa \mu}(r) S_{\kappa \mu}(\theta) ,
\end{equation}
which is calculated as
\begin{equation}
\left( r^{2}R_{\kappa \mu}(r), S_{\kappa \mu}(\theta) \right) \! = \!
\left( 2G_{\kappa \mu}^{(0)}(r)F_{\kappa \mu}^{(0)}(r) ,
\, -\!\!\!\sum_{L={\rm odd}} \!\!\!
 \xi_{\kappa \tilde{\kappa}}^{LL}C_{\kappa \tilde{\kappa} \, \mu}^{LL} i {\cal Y}_{L \,1}(\theta \, 0)
 \right).
\end{equation} 
The radial wave functions $G_{\kappa \mu}^{(0)}(r)$ and $F_{\kappa \mu}^{(0)}(r)$ are
depicted in l.h.s. of Fig. \ref{crntfld}.
The cross section of convection like current $q_{\kappa \mu}(r \theta)$, the radial form factor
$R_{\kappa \mu}(r)$ and angular form factor $S_{\kappa \mu}(\theta)$ are shown in r.h.s. of Fig. \ref{crntfld}
\footnote{The decomposition of currents into convection and spin currents was defined by Gordon only for
free particle.  Since the terms used here is not exact, we say convection or spin "like" currents.} 
\subsection{Circular Helicity in Weyl Equation}
\label{crhlcyW}
The symmetry breaking of chirality is brought about by the electron capture through parity
violating weak interaction. The negative chirality state is absorbed and the positive state remains.
We will demonstrate the predominance of positive helicity states in the positive chiral state by means of
a simple numerical calculation in this subsection.  
 \begin{figure}\label{predomi}
\centerline{\includegraphics[width=14 cm,height=4 cm]{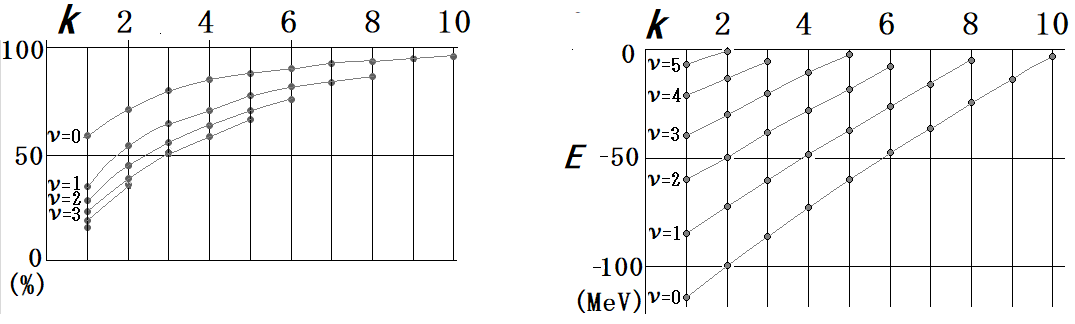}}
 \caption{Pulity of circular helicity $h_{k\nu}$of the positive chirality states is plotted in l.h.s. versus
 $k=\mid \kappa \mid$ with connecting same radial node number $\nu$. The energy
 spectra are plotted in r.h.s. }
 \end{figure}
With neglecting mass term, Weyl equation for the two components $H^{(c)}_{h j \mu}$ 
in eigenvector of chirality $c$ is given
\begin{equation}
\label{Weyleq}
\left( \!\!\! \begin{array}{cc}  \displaystyle (-1)^{c-h} \frac{k}{r} +u(r) &\displaystyle (-1)^{c-h}\frac{d}{dr}
\vspace{0.2cm} \\ \hspace{-0.3cm}
\displaystyle -(-1)^{c-h}\frac{d}{dr} & \hspace{-0.6cm} \displaystyle -(-1)^{c-h} \frac{k}{r} +u(r) 
\end{array} \!\!\! \right) \!\!
\left( \!\! \begin{array}{c} H^{(c)}_{\!\frac{1}{2}j\mu} (r)
\vspace{0.1cm} \\ H^{(c)}_{\! \frac{\!-\!1}{2}j\mu} (r) \end{array} \!\!\right)=
\varepsilon
\left( \!\! \begin{array}{c} H^{(c)}_{\frac{1}{2}j\mu} (r)
\vspace{0.1cm}\\ H^{(c)}_{\!\frac{\!-\!1}{2}j\mu} (r) \end{array} \!\! \right) \, ,
\end{equation}
where $c=\pm \frac{1}{2}$ expresses positive and negative chirality, respectively.
The differentiation operator with respect to radial coordinate is in off-diagonal
and therefore mixing of circular helicity is taken place through radial motion. 
It results reduction of purity of circular helicity for large radial node
numbers $\nu$.
We solve eq. (\ref{Weyleq}), by using the same potential $u(r)$ and technique
obtaining quasi-bound states, to obtain the stationary states.
The purity of helicity is defined by
\begin{equation}
 h_{\kappa \nu}=\frac{ {\displaystyle \int_{0}^{\infty}}{dr \left[ H_{\frac{1}{2} j \mu}^{(\frac{1}{2})}(r)^{2} 
 -H_{\!\frac{\!-\!1}{2} j \mu}^{(\frac{1}{2})}(r)^{2}\right]}}{ {\displaystyle \int_{0}^{\infty}}dr \left[
  H_{\frac{1}{2} j \mu}^{(\frac{1}{2})}(r)^{2} +H_{\!\frac{\!-\!1}{2} j \mu}^{(\frac{1}{2})}(r)^{2} \right] }\, .
 \end{equation}
Fig. \ref{predomi} clearly shows the predominance of probability of state having the same sign of circular helicity as chirality 
in the higher $k$ with the less radial node number.
Particularly the orbital relevant to forming the toroidal MF, namely the highest $k(=10)$ , the highest $\mu=\frac{19}{2}$
and $\nu=0$, possesses the highest pulity.
\subsection{Helical Current}\label{HlclCrrnt}
  In Weyl representation, current density is calculated separately plus and minus chirality like
 in eq.(\ref{dncrdn_W}).
 The remnant of the reaction of electron capture is composed of positive chirality,
so that neglecting mixing of chirality, we calculate the current using the wave function obtained
in the former subsection.

A helicity representation state includes two states of different sign of $\kappa$. 
Hence, the probability current between different sign of $\kappa$ states gives rise to
magnetic odd moments, i.e. convection like circular current as shown in subsection
\ref{ECuRNS} . 

The diagonal current of eigenstate of chirality are decomposed into two terms,
namely, spin and
convection like currents, such as \vspace{0.1cm}
\begin{equation}
\vect{j}_{k}(r,\theta)=\vect{j}_{k}^{({\rm sp})}(r,\theta)+\vect{j}_{k}^{({\rm cv})}(r,\theta),
\end{equation}
which are expressed by
\begin{equation}
\left\{
\begin{array}{l} \displaystyle
\vect{j}_{k}^{({\rm sp})}(r,\theta)= c_{0} \sum_{i=1}^{2} R_{i}(r)
\left(S^{(i)}_{\rho}(\theta)\vect{n}_{\rho} +S^{(i)}_{z}(\theta)\vect{n}_{z} \right)
\vspace{0.2cm}\\
\vect{j}_{k}^{({\rm cv})}(r,\theta)= c_{0} R_{3}(r)S^{(3)}_{\phi}(\theta)\vect{n}_{\phi} 
\end{array}
\right. ,
\end{equation}
\begin{figure}
\centerline{\includegraphics[width=10 cm,height=5 cm]{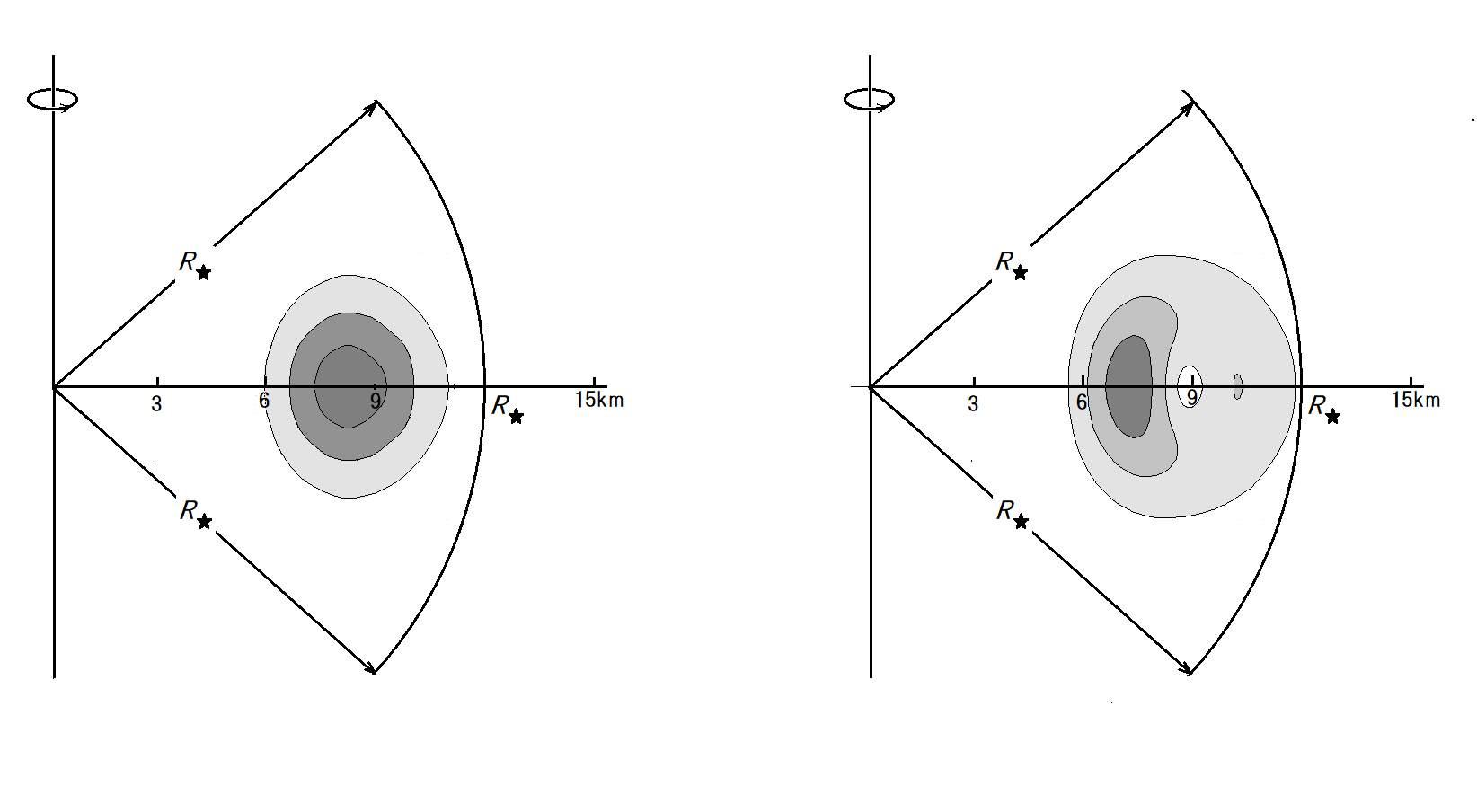}}
\caption{Density distribution of cross section of convection like current is shown in l.h.s. and
density distribution of spin like current is depicted in r.h.s..}
\label{cnvspncrr}
\end{figure}
where three radial form factors $R_{i}(r)$ are calculated by using two helicity component
of wave functions $(H_{\frac{1}{2}}(r) \, H_{\!-\!\frac{1}{2}}(r))$, omitting the superscript $(c)\!=\!(\frac{1}{2})$
and suffixes $j\!=\!\frac{19}{2}$ and $\mu\!=\!\frac{19}{2}$, as follows
\begin{equation}
\left( \begin{array}{c}
R_{1}(r),R_{2}(r) \vspace{0.15cm} \\
R_{3}(r)
\end{array} \right)=\frac{1}{r^{2}}
\left( \begin{array}{c}
H_{\frac{1}{2}}^{2}(r)\!\!+\!\!H_{\!\frac{\!-\!1}{2}}^{2}(r), \,
2 H_{\frac{1}{2}}(r)H_{\!\frac{\!-\!1}{2}}(r), 
\vspace{0.15cm} \\
 H_{\frac{1}{2}}^{2}(r)\!\!-\!\!H_{\!\frac{\!-\!1}{2}}^{2}(r)
 \end{array}
\right) ,
\end{equation}
and angular dependent factors are calculated like
\begin{equation}
\left( \begin{array}{c} 
S^{(i)}_{\rho}  (\theta) \vspace{0.15cm} \\
S^{(i)}_{z}  (\theta) 
\end{array} \right)
=\sum_{L} \cos{\left(\frac{L\pi}{2}\right)}
\left( \begin{array}{l}
S^{(i)}_{\rho \, L} {\cal Y}_{L \, 1}(\theta \, 0) \vspace{0.15cm} \\ 
S^{(i)}_{z \, L} {\cal Y}_{L \, 0}(\theta \, 0)
\end{array} \right) ,
\end{equation}
from following four coefficients
\begin{equation}
\left( \begin{array}{cc} 
S^{(1)}_{\rho \, L} & S^{(2)}_{\rho \, L} \vspace{0.15cm} \\
S^{(1)}_{z \, L}  & S^{(2)}_{z \, L} 
\end{array} \right)
=\sum_{J=L\pm 1} C^{J}_{k}
\left( \begin{array}{cc}
\pm \hat{J}C_{J}^{(\pm)}C_{L}^{(\pm)} & 2k C_{J}^{(\mp)}C_{L}^{(\pm)}
\vspace{0.15cm} \\
\pm \hat{J}C_{J}^{(\pm)}C_{L}^{(\mp)} & 2k C_{J}^{(\mp)}C_{L}^{(\mp)}
\end{array} \right) ,
\end{equation}
with
\begin{equation}
C^{J}_{k}=\langle j j j -j\mid J 0 \rangle d^{J}_{k k}\hat{J}^{-1} .
\end{equation}
The spin like current is parallel to meridian plane having vector components
$\vect{n}_{\rho}$ and $ \vect{n}_{z}$ only, because of vector spherical
harmonics for $L=J\pm 1$ and $M=0$.  
The angular dependent factor of convection like current is given as
\begin{equation} 
S_{\phi}^{(3)}(\theta)=2k\sum_{L} \cos{\frac{(L\!-\!1)\pi}{2}}
C^{L}_{k}(-i{\cal Y}_{L \, 1}(\theta\,0))\, .
\end{equation}
 \begin{figure}
 \label{spincrrnt}
\centerline{\includegraphics[width=11 cm,height=4.7 cm]{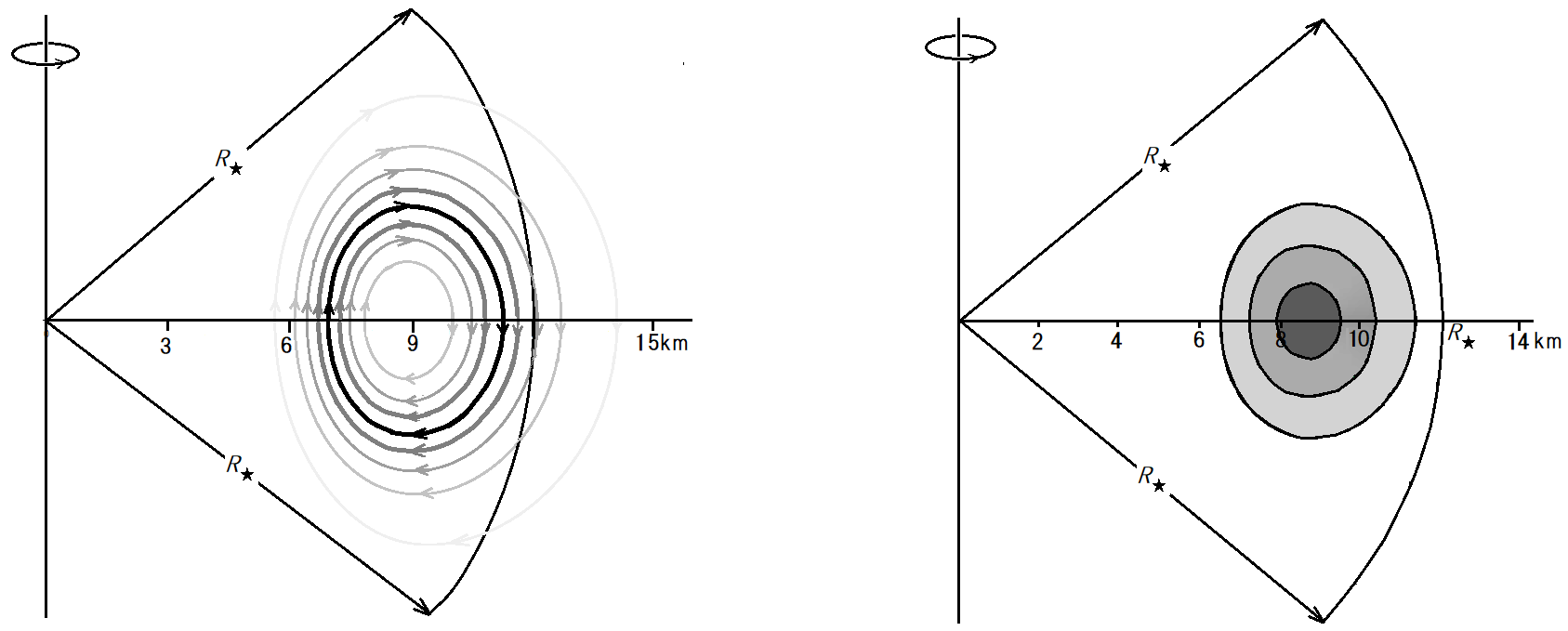}}
 \caption{ The field lines of spin like current on a meridional plane are depicted in l.h.s.. 
 The contour map of cross section for strength of toroidal MF is shown in r.h.s. .}
 \end{figure}
Eventually, the convection like current is perpendicular to meridional plane,
while the spin like current is parallel to the plane.
So that, two kinds of current are orthogonal to each other and form helical current.
This helical current is very characteristics resulted from the Dirac current in asymmetric
chirality states given rise to by parity violating weak interaction.  
In the case of axial symmetry, the MF, therefore  is perpendicular to the plane,
having only component of $\phi$ which is function of $(r, \theta)$, i.e., $B_{\phi}(r,\theta)$ 
and is calculated, conveniently to use of cylindrical coordinates $(\rho=r\sin \theta, \phi, z=r \cos \theta)$ as
\begin{equation}
B_{\phi}(\rho,z)=\frac{e\mu_{0}}{2\pi \rho} \int_{0}^{\rho}
 j_{z}(\rho^{\prime}, z) (2\pi \rho^{\prime}) d\rho^{\prime} ,
 \end{equation}
by Ampere's law.
The maximum strength occurs at circle of radius 9 km on the equatorial plane
( see Fig. \ref{spincrrnt}), and the
strength turns out to be approximately $3.2 \gamma^{3}  \times 10^{16}$ TT.
We calculate for five points of scaling factor $\gamma=2^{-n}  (n=3,5,7,9,11)$ and
tabularize the results in Table \ref{trdlfld}.  
These large numbers come from the factor
$e\mu_{0}c_{0}N_{\star}/R_{\star}^{2}=4.19\times 10^{17}$TT.
The $\gamma^{3}$ dependence corresponds to size of collectivity 
$N_{\ast}=\gamma^{3} N_{\star}$. 
%
%
\subsection{Magnetic Helicity}
\label{MgnHel}
 The magnetic helicity is expressed by the volume integral of
 \begin{equation}
 h_{\rm m}=\int_{V}  \vect{A}(\vect{r}) \cdot  \vect{B}(\vect{r}) d \vect{r} ,
\hspace{0.5cm} {\rm with} \hspace{0.5cm}
 \vect{B}(\vect{r}) = \vect{\nabla} \times \vect{A}(\vect{r}).
 \end{equation}
 The operator $\nabla$ changes the parity of the operand field and therefore only the integral is nonzero for
 the cross terms of even and odd parity fields.
It was shown in \ref{HlclCrrnt} that, in the axially symmetric configuration,
the convection like current has only the component of $\vect{n}_{\phi}$ and
 the MFs parallel to the meridian plane
 while the spin like current and its vector potential are parallel to the meridian plane
 and the MF has only the component of $\vect{n}_\phi$.
 Eventually, the magnetic helicity is contributed from the cross terms of the convection like current and 
 the spin like current as follows
  \begin{equation}
  \begin{array}{r} \displaystyle
  h_{\rm m} \!\! = \!\!
\int_{V} \!\! \left( \! \vect{A}^{\rm (sp)}\!(\vect{r}) \!\cdot \!\! \vect{B}^{\rm (cv)}(\vect{r})
\!+\! \vect{A}^{\rm (cv)}(\vect{r}) \!\cdot\!\!  \vect{B}^{\rm (sp)}(\vect{r}) \right)d^{3} \vect{r}
\!\!=2\pi \!\! \int_{0}^{\infty} \!\!\! r^{2} dr  \!\! \int_{0}^{\pi} \!\!\!  \sin \theta d\theta
\vspace{0.3cm} \\ \displaystyle \hspace{0.5cm} \times
\left( \! A_{r}^{\rm (sp)}(r \theta) \! \cdot \! B_{r}^{\rm (cv)}(r  \theta)
\! + \! A_{\theta}^{\rm (sp)}(r \theta) \! \cdot \! B_{\theta}^{\rm (cv)}(r \theta)
\! + \! A_{\phi}^{\rm (cv)}(r \theta) \! \cdot \! B_{\phi}^{\rm (sp)}(r \theta)
\right).
\end{array}
 \end{equation} 
 The positive chirality states dominate positive helicity and have positive magnetic helicity. 
This is resulted from the fact that neutrino has negative chirality in nature. Hence the magnetic helicity
is not a topological invariance but the nature of Dirac (Weyl) helical current described by Clifford number.
\subsection{Selective Capture and Sustainability of Chiral Asymmetry}
\label{ChAsHelCrr}
Now, we come to the final stage of the scenario for the mechanism forming
the strong toroidal MF.
The helical current composed from convection like current caused by orbiting electrons
and spin like current generated by alignment of electron spin pointing to tangential of the circular orbit.
The orbiting  electrons are captured by protons through parity violating weak interaction.
While only electrons in negative chiral state selectively involve in the reaction,
electrons having positive chirality remain in the orbiting zone.
This results the chiral asymmetry, 
and therefor remaining electrons possess dominantly positive helicity.
The positive helical current gives rise to positive magnetic helicity,
because of the similarity relation (\ref{Green_fn})
between the moments of current and vector potential. 

In the absence of vector potential $\vect{A}(r)$, the energy eigenvalue of positive chirality states is exactly same as one of corresponding negative chirality states, namely pairwise degenerate.
Therefore, the mass term being off-diagonal in the Weyl representation mixes the two states with equal weights, so that resulting wave functions have definite signature of $\kappa$, i.e. the canonical states through the inverse transformation of eq. (\ref{cir_hel}) to make the mass term diagonal.
It will show the chiral asymmetry developing under the MF produced by the helical currents of plus chirality state. Without changing each wave function for the different chirality states, the matrix elements of interaction
\begin{equation}
\langle \psi^{(c)} (\vect{r} \sigma)\mid c_{0}\vect{\sigma} \cdot \vect{A} \mid 
\psi^{(\frac{1}{2})}(\vect{r} \sigma)\rangle_{\sigma}=-e^{2}c_{0}^{2}\mu_{0}N_{*}(-1)^{c-\frac{1}{2}}\sum_{LJM}v^{(c)}_{LJM}\, ,\vspace{-0.3cm}
\end{equation}
is calculated by a double integral
\begin{equation}
v^{(c)}_{LJM}=\frac{(-1)^{J-M-1}}{2L+1}\!\!\int_{0}^{\infty}\!\!r^{2}dr \int_{0}^{\infty} \!\!\! r^{\prime \, 2}dr^{\prime}
\eta_{LJ\,-\! M}^{(c)}(r)\left( \frac{r^{L}_{<}}{r^{L+1}_{>}}\right)\eta_{LJM}^{(\frac{1}{2})}(r^{\prime})\, .
\end{equation}
In the present case, the integral is expressed in a separate form
\begin{equation}
\sum_{LJ}v^{(c)}_{LJ0}=
\sum_{L={\rm even}}\sum_{J=L\pm 1}(C^{J}_{k})^{2} n_{LJ}^{(c;{\rm \,sp)}}+
\sum_{L=J={\rm odd}}(C^{J}_{k})^{2}n_{L}^{(c;{\rm \,cv)}}\, ,
\vspace{-0.3cm}
\end{equation}
where
\vspace{-0.1cm}
\begin{equation}
\left( \begin{array}{c}\!\!\!
n_{L\, J=L\pm 1}^{(\!c;{\rm \,sp)\!} } \vspace{0.1cm} \\  n_{L}^{(\!c;{\rm \,cv})} 
\end{array} \!\!\! \right)\!
=\!\frac{1}{2L\!+\!1}\! \int_{0}^{\infty}\!\!\!r^{2}dr \!\int_{0}^{\infty} \!\!\!\! r^{\prime 2}dr^{\prime}
\left( \begin{array}{c} \!\!R_{\pm}^{(c)}(r) \vspace{0.1cm}\\ R_{3}^{(c)}(r) 
\end{array} \!\!\right)
\left( \frac{r^{L}_{<}}{r^{L+1}_{>}}\right)
\left( \begin{array}{c} \!\!R_{\pm}^{(\!\frac{1}{2}\!)}(r)\vspace{0.1cm} \\ R_{3}^{(\!\frac{1}{2}\!)}(r)
\end{array} \!\!\right) \, ,
\vspace{-0.2cm}
\end{equation}
with
$
R_{\pm}^{(c)}(r)=\pm \hat{J}R_{1}^{(c)}(r)+2k\,R_{2}^{(c)}(r) \, .
$
From the symmetry character of Hamiltonian expressed in eq.(\ref{Weyleq}), 
the wave functions have correspondence
$H_{-h}^{(-c)}=H_{h}^{(c)}$, and therefore $R_{\pm}^{(c)}=R_{\pm}^{(-c)}$ and
$R_{3}^{(c)}=-R_{3}^{(-c)}$ or $n_{LJ}^{(c;{\rm \,sp)}}=n_{LJ}^{(-c;{\rm \,sp)}}$ and 
$n_{L}^{(c;{\rm \,cv)}}=-n_{L}^{(-c;{\rm \,cv)}}$.
With taking into account of the relations
$\vect{j}^{\pm}=\langle \psi^{\pm} \mid \pm c_{0} \vect{\sigma}\mid\psi^{\pm}\rangle$ in eq. (\ref{chrlcrrnt}), 
only spin like current changes its sign.
Eventually, the diagonal matrix element of the vector potential works attractive for the positive chirality state
and repulsive for the negative chirality state and therefore pushes up the energy of negative chirality state
and vice versa, immediately after the reaction within microsecond, i.e.  time of MF propagating.
Numerical results of the energy
 $v\!\!=\!\! \langle \psi \!\!\mid \! c_{0} \vect{\sigma}\! \cdot \!\vect{A}^{({\rm sp})} \!\mid \!\! \psi \rangle_{\sigma}$
are tabulated in Table \ref{trdlfld}. 
This value depends on $\gamma^{4}$,
because the cross-section area of the toroidal MF is proportional to $\gamma$; 
elliptical minor axis $r_{\rm t} \propto \sqrt{\gamma}$ .
In the cace of 8 sec rotation period, the value is estimated as $\approx 1 \times 10^{18}$MeV, which is much larger
then electron mass $m_{\rm e} c_{0}^{2} = 0.51$MeV, being negligible small. 
These big values come from the big value $e\mu_{0} c_{0}^{2}N_{\star}/R_{\star}=1.5 \times 10^{36}$MV.
\begin{table}[htbp]
\begin{center}
\begin{tabular}{|c|ccccc|}
\hline
$\gamma=2^{-n}$ & $n=3$ & $n=5$ & $n=7$ & $n=9$ & $n=11$  \\
\hline
 $k$    &  43  & 176  & 708 & 2838 & 11355 \\
 $\Delta E$ (MeV) & 2.69 & 0.675 & 0.169 & 0.0423 & 0.0106 \\
 $P$(msec) & 1.65 & 67.3 & 270 & 1085 & 4340 \\
 $B_{\rm max }$ (TT) & 5.16 $\times 10^{13}$ & 9.60 $\times 10^{10}$ &
1.56$\times 10^{10}$ & 2.50 $\times 10^{8}$ & 3.96 $\times 10^{6}$\\
  $v$ (MeV) & 8.40$\times 10^{28}$ & 2.97 $\times 10^{26}$ & 1.13 $\times 10^{24}$
& 4.36 $\times 10^{21}$ & 1.70 $\times 10^{19}$ \\
  $r_{\rm t}$ (km) & 1.061 & 0.530 & 0.265 & 0.133 & 0.066 \\
 \hline  
 \end{tabular}
 \end{center}
\caption{Scale factor $\gamma$ dependence of the variables, quantum number; $k=j+1/2$, 
difference of energy between 
unoccupied and occupied states; $\Delta E= E_{k+1}-E_{k}$, rotation period; $P$,
maximum strength of toroidal MF; $B_{\rm max}$, diagonal matrix element of vector potential of spin like current;
$v$, and size of intersection of toroidal MF, are listed respectively.} 
\label{trdlfld}
\vspace{-0.2cm}
\end{table}

In conclusion, the negative chirality state gains energy from repulsive force available to electron capture,
while the positive chirality state loses energy from attractive force to become stable. 
The degeneracy with positive and negative chirality states is solved to accelerate the chiral asymmetry reaction
and, at the same time, to sustain the chiral asymmetry against decay of the state due to the mass term.
In other words, the toroidal MF produced by the spin like current supports the current itself.
\section{Discussions} 
In this work, we have theoretically studied the origin of strong toroidal magnetic field in a typical size of magnetar of $M=1.5 M_{\odot}$ and $R=12$km. It is realized that the electrons as an entire system is in quantum degenerate state with ultra-relativistic Fermi energy. We analyzed the system with Dirac Hartree-Fock like theory with scaled h-bar method, to find that the spin like current plays a crucial role in forming the toroidal MF.
 We employed the cranking model which is used in investigating collective rotational motion of quantum many-body systems such as nucleus or Bose-Einstein condensates.
The cranking model is characterized by the so called Coriolis term $-\hat{\vect{J}} \omega $ where the angular velocity $\omega$ works as Lagrange multiplier in energy minimization together with conservation of total AM. The cranking term $-\hbar_{*} \mu \omega$ breaking time reversal, splits energy of single quantum state proportional to the magnetic quantum number $\mu$. Hence the energy of quantum state $j$ having the lowest magnetic quantum number, $\mu =-j$ is the highest and vice versa. At the level crossing of the two states, the highest $\mu=-j$ of occupied state and the lowest $\mu^{\prime}=j+1$ of unoccupied state, the particle-hole like excitation causes a collective rotation because of asymmetric occupation of time reversal conjugate pairs $\pm \mu$ and constructs the convection like current flax in peripheral region near equatorial plane.

The selective electron capture through parity violating weak interaction leaves positive chirality electrons having dominantly positive circular helicity as remnants resulting in chiral asymmetry population. The magnetic helicity is determined by the relative phase between convection like and spin like currents. We calculated the global structure of both currents for a chiral eigen-state and found that the convection like current is formed by magnetic odd moments of vector field and the spin like current is formed by electric odd moments, respectively. The former produces the electric odd moments of MF to be a poloidal structure and the latter magnetic odd moments of MF to be a toroidal, respectively. Eventually, we have succeeded to explain the origin of the toroidal MF based on the spin like current caused by Dirac current being Clifford number.

We calculated the diagonal matrix elements for the scaler product of each current and vector potentials of currents of positive chirality eigenstate. It is seen that the spin like current of positive chirality state acts as attractive force and that of negative chirality state acts as repulsive force. This force solves degeneracy of the energies of quantum states and pushes up the negative chirality state and pushes down the positive chirality state. This effect works as acceleration, at the same time, as sustainability of chiral asymmetry against mixing through mass term, which is the off-diagonal element in the Weyl representation.
We proved numerically the diagonal matrix element to be much larger then the electron mass. 
As for the convection current, contribution from protons is competitive with that of electrons, while contribution of toroidal MF from protons are negligible due to the non-relativistic feature coming from 1840 times larger mass than that of electrons.

 One limitation in this paper is that we did not consider the hadronic system in detail.
 This shall be studied in a different paper. The proton circular current acts repulsive to the convection like current of electrons. As a result, two polar axes slightly split in balance with the increase of additional rotational energy caused by the splitting. This splitting gives rise to precession motion, which may explain the phase modulation of hard X-ray from soft X-ray period observed by Makishima et. al. \cite{ MEH14} .
A time dependent theoretical calculation may be necessary to explain increase of chiral asymmetry due to electron capture and stabilization of the toroidal MF due to helical current.

Another limitation is that we completely ignored the general relativity which may play an important role.
In order to tackle this limitation, the metric tensor has to be obtained possibly by using 
TOV's equation, and Maxwell equation has to be solved in the curved space.  
%
%

\vspace{0.4cm}
\hspace*{3.5cm} {\Large { Acknowledgement}}

The authors express hearty thanks to Prof. K. Makishima for suggesting the present subject
and encouraging us to work out these results.
We are also indebted to Profs A. Ohnishi and N. Yamamoto for giving a clue
to solve the problem in terms of chiral asymmetry and magnetic helicity, 
to Professor K. Iida for making clear the electric monopole polarization and confinement,
to Dr. H. Sotani  for illuminating discussions on time dependent theory in forming the MF,
and to Profs. O. Yasuda and T. Hyodo
for use of facilities at TMU.
This work is supported by Grants-Aid for the Scientific Research from the Ministry of Education Science and
Culture of Japan (16K05360).

\end{document}